\documentclass[a4paper,11pt]{article}

\usepackage{jinstpub}

\usepackage{graphicx}%
\usepackage{multirow}%
\usepackage{amsmath}%
\usepackage{amsthm}%
\usepackage{mathrsfs}%
\usepackage[title]{appendix}%
\usepackage{xcolor}%
\usepackage{textcomp}%
\usepackage{array}%
\usepackage{manyfoot}%
\usepackage{booktabs}%
\usepackage{algorithm}%
\usepackage{algorithmicx}%
\usepackage{algpseudocode}%
\usepackage{listings}%
\usepackage{subcaption}%
\usepackage{siunitx}%
\usepackage[percent]{overpic}%
\usepackage{comment}

\begin{document}

\title{Photoluminescence of resin-based solder flux residue under ultraviolet excitation from 120 nm to 310 nm}

\author[a,b]{Anna Hurhina}
\author[b,1]{Marjolein Troost\note{Corresponding author.}}
\author[c]{Andreas Leonhardt}
\author[a,b]{Tina R. Pollmann}

\affiliation[a]{Institute of Physics, University of Amsterdam,\\
1098 XH Amsterdam, Netherlands}

\affiliation[b]{Dark Matter Group, Nikhef,\\
1098 XG Amsterdam, Netherlands}

\affiliation[c]{Department of Physics, TUM School of Natural Sciences,
Technical University of Munich,\\
85748 Garching b. Munich, Germany}

\emailAdd{mnuland@nikhef.nl, t.pollmann@nikhef.nl}

\abstract{Nuisance photoluminescence is a potential source of background in particle detectors that use noble liquids as target material for galactic dark matter particles and neutrinos. Liquid argon and xenon scintillate in the vacuum ultraviolet (VUV) wavelength range in response to particle interactions. Photoluminescent materials that absorb these photons can cause unexpected signals that may impede event reconstruction in these detectors. We illuminated residue from different types of commercial solder flux commonly used in liquid xenon detectors with ultraviolet and VUV light and measured their photoluminescence spectra and intensities. We find that all tested flux residues photoluminesce in the visible spectral region when exposed to VUV light.}


\maketitle
\arxivnumber{2605.16845} 

\section{Introduction}\label{sec:introduction}
Some of the most sensitive particle detectors that study neutrinos and dark matter use liquid argon (LAr) or liquid xenon (LXe) as interaction targets~\cite{ds-50,detectorpaper,aalbersFirstDarkMatter2022,aprileXENONnTDarkMatter2024,DS20K,abdukerimPandaXxTADeepUnderground2024,collaborationXLZDDesignBook2024}. When a particle interacts with the noble liquid, vacuum ultraviolet (VUV) scintillation light is produced. The number and arrival times of the individual scintillation photons are measured and used to reconstruct the energy, location and nature of the interaction.
Detector sensitivity can be significantly degraded if scintillation photons excite photoluminescent contaminants within the detector. Delayed photon emission from photoluminescence (PL) modifies both the timing and the total number of detected photons, potentially leading to misreconstruction. For this reason, nuisance PL has been discussed as a possible source of unexplained backgrounds in several large noble-liquid experiments~\cite{akeribEnhancingSensitivityLUXZEPLIN2021, sorensenTwoDistinctComponents2018, akerib_investigation_2020, aprile_emission_2022}. Since these detectors trigger on as little as 2-4~detected photons, and some interactions produce hundreds of thousands of scintillation photons, PL efficiencies\footnote{Defined as the number of PL photons emitted relative to the number of UV photons incident on the PL impurity.} below 1\% can already cause significant detector backgrounds depending on the location of the contaminant~\cite{singh_search_2025}.

Most LAr and LXe detectors operate as dual-phase time projection chambers (TPCs), in which an active volume of noble liquid is defined by a hollow polytetrafluoroethylene (PTFE) cylinder and bounded at the top and bottom by electrode grids that establish a uniform electric drift field along the cylinder axis. Scintillation light is produced in both the liquid and gas phases and, while predominantly in the vacuum ultraviolet (VUV) range (peaked at \SI{128}{\nano\meter} for LAr~\cite{Argon_scintillation_wavelength} and \SI{175}{\nano\meter} for LXe~\cite{Xe_wavelength}, both with approximately \SI{10}{\nano\meter} full width at half maximum), up to $\sim$10\% of photons are emitted at longer wavelengths from \SIrange{250}{400}{\nano\meter}~\cite{bakshtSpectralCharacteristicsHighcurrent2007, condeGasProportionalScintillation1968,aoyamaMeasurementEmissionSpectrum2022a}. The TPCs are operated inside cryostats that keep the target material at temperatures of approximately \SI{87}{K} (LAr) or \SI{175}{K} (LXe)~\cite{XENON1T2017, DS20K}.

The photons are detected by arrays of photomultiplier tubes (PMTs) with sensitivity across the VUV-IR regime, located above and below the active volume. Numerous electrical components, including PMT bases, temperature sensors, and level meters, are directly immersed in the noble liquid and are assembled using soldered connections~\cite{aprileXENONnTDarkMatter2024,xenon_collaboration_design_2024}.

 Contaminants that exhibit PL at the relevant intensities and/or timescales have not been conclusively identified in these detectors. The main materials in contact with the noble liquid volumes - PTFE and the PMT windows - and possible contaminants on or in them, have been investigated with inconclusive results\cite{sorensen_two_2018,araujoPhotoluminescenceResponseAcrylic2019,sorensen_quartz_2025}.
 
The dark matter detectors DarkSide-50~\cite{collaborationElectronicsTriggerData2017} (a LAr-based TPC), XENON1T, XENONnT~\cite{brownSearchElasticInelastic2020}, and LUX-ZEPLIN (LZ)~\cite{akeribLUXZEPLINLZRadioactivity2020} (LXe-based TPCs) use solder fluxes with rosin (RO) or resin (RE) classification, according to industry standard J-STD-004~\cite{RequirementsSolderingFluxes1995}. These fluxes contain alcohols and organic acids, such as abietic acid (C$_{20}$H$_{30}$O$_2$), which, when heated during production or soldering, can form photoluminescent compounds such as polycyclic aromatic hydrocarbons (PAH) that are left behind in the flux residue.

While common resin-based solder fluxes are known to photoluminesce under UV-A (\SIrange{315}{400}{\nano\meter}) excitation~\cite{spectrum_and_cleaning_procedure}, and this property is routinely used in electronics manufacturing, photoluminescence efficiency and emission spectra of these residues have not been characterised. We study the PL of the residue from solder fluxes used in the XENONnT and LZ detectors at excitation wavelengths and sample temperatures relevant to LAr and LXe-based dark matter detectors.

\section{Spectrofluorometry of solder  residue}\label{sec:spectrofluorometry}

Ideally, one would measure the PL spectra from each sample at LAr and LXe temperature in a setup with well-calibrated excitation photon flux at LAr and LXe wavelengths. Due to the experimental difficulties of combining the detection of small numbers of PL photons (because of low PL efficiencies) with cryogenic temperatures, and VUV vacuum and light systems, we performed measurements in two separate setups. The first setup was used to measure PL emission spectra under a fixed wavelength of \SI{310}{\nano\meter} excitation light as a function of sample temperature. From this, we obtain the relative change in PL spectra and efficiency between room temperature and LAr/LXe temperature. We then used the second setup to measure wavelength-integrated PL efficiencies at room temperature and as a function of excitation wavelength in the VUV regime, and compare results to a reference sample to obtain absolute efficiencies. By combining the two results, we then estimate the PL efficiency (relative to the reference sample) of the solder flux residue at the temperatures and excitation wavelengths of interest.

\subsection{Sample preparation}
We prepared four samples: solder flux residue from flux used in the XENONnT detector, solder flux residue from flux used in the LZ detector, a PL reference sample, and a non-photoluminescent blank. The four samples were prepared for this study in the following way:
\begin{enumerate}
\item \textbf{Blank:} Three copper plates of dimensions \SI{0.1}{\centi\meter}$\times$\SI{3}{\centi\meter}$\times$\SI{3}{\centi\meter} were ultrasonically cleaned. One plate was used as a blank, while the remaining plates served as substrates for the flux samples. Copper was chosen as the blank and carrier material because it does not photoluminesce and provides good thermal conductivity.
\item \textbf{Flux residue 1} (Stannol KS115, resin-based flux containing halides, classified REM1~\cite{KS115_datasheet}, used in the XENONnT detector): A flux-cored solder wire was melted onto the copper substrate using a soldering iron set to \SI{380}{\degree}C. While still molten, the solder alloy was removed, leaving only the flux residue on the substrate.
\item \textbf{Flux residue 2} (Chemtronics CW8400, rosin-based flux, classified ROL0~\cite{CW8400_datasheet}, used in the LZ detector): The flux, supplied in a dispenser pen, was deposited directly onto the copper substrate and subsequently heated to \SI{380}{\degree}C to reproduce soldering conditions.
\item \textbf{Reference:} A plate of detector-grade polyethylene naphthalate (PEN) with dimensions \SI{0.5}{\centi\meter} $\times$ \SI{3}{\centi\meter} $\times$ \SI{3}{\centi\meter}, produced by the LEGEND collaboration~\cite{manzanillas_optical_2022}, was used as a reference. PEN is a photoluminescent polymer with known optical and PL properties~\cite{araujoWavelengthshiftingReflectorsCharacterization2022,boulayDirectComparisonPEN2021}.
\end{enumerate}
Photographs of the samples under white LED and UV-A (\SI{365}{\nano\meter}) illumination are shown in Fig.~\ref{fig:sample_photos}. To minimise contamination by VUV-absorbing or photoluminescent organic impurities, all samples were handled with nitrile lab gloves and while wearing face coverings, and stored in individual sealed containers between use. The copper and PEN samples were cleaned with lint-free wipes soaked in isopropanol prior to each measurement.

\newcommand{\scalebarimage}[1]{%
\begin{overpic}[width=0.3\textwidth]{#1}
    \put(60,8){\color{white}\makebox[20mm][l]{\rule{20mm}{1.5pt}}}
    
    \put(72,12){\color{white}\small $\sim$ \SI{2}{\centi\meter}}
\end{overpic}
}

\begin{figure}[!htb]
\centering
\begin{tabular}{m{1.5cm}m{0.3\textwidth}m{0.3\textwidth}}
 Sample & Under white LED light & Under UV light (\SI{365}{\nano\meter})\\
 KS115 &
 \scalebarimage{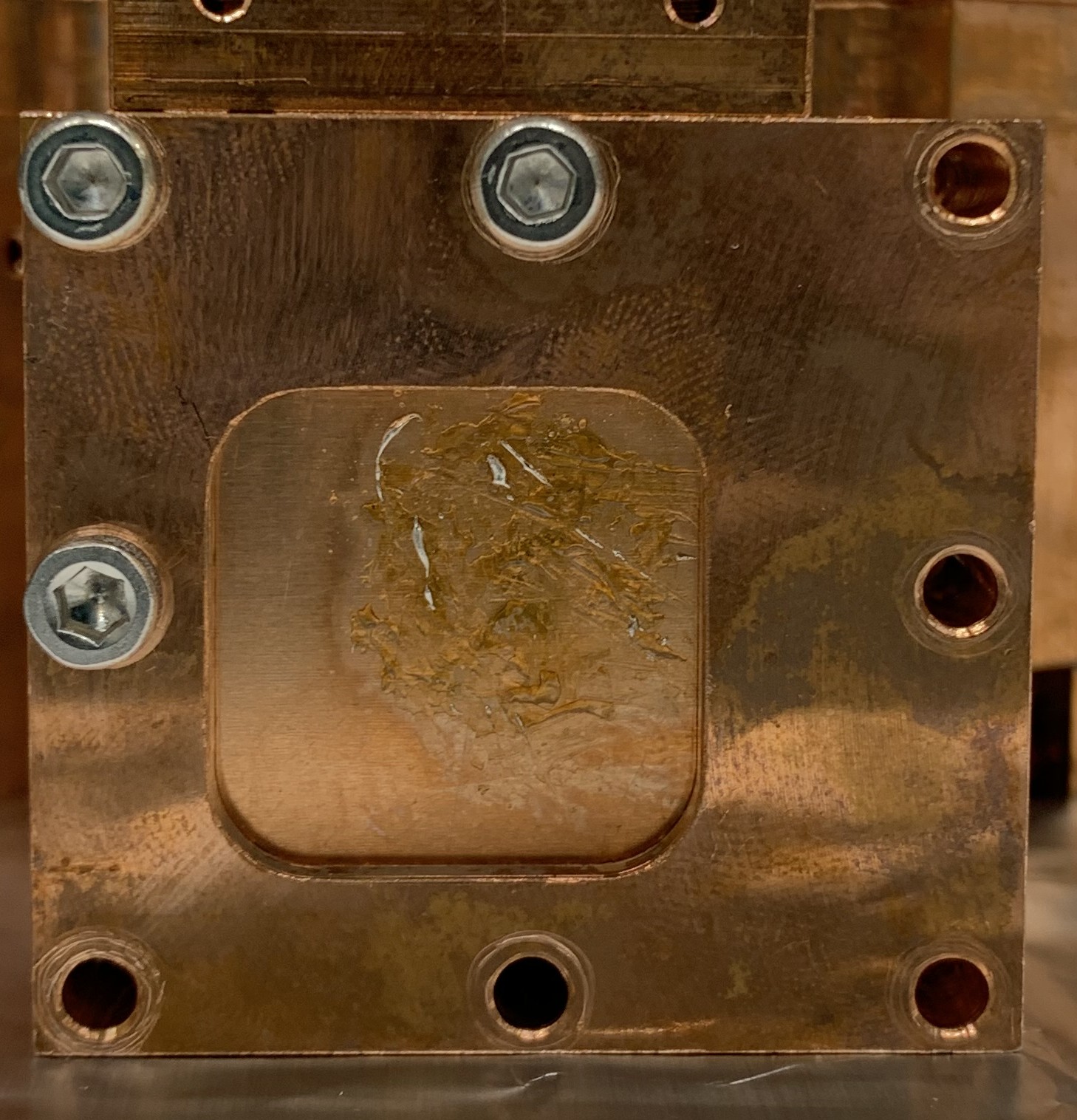}
 &
 \scalebarimage{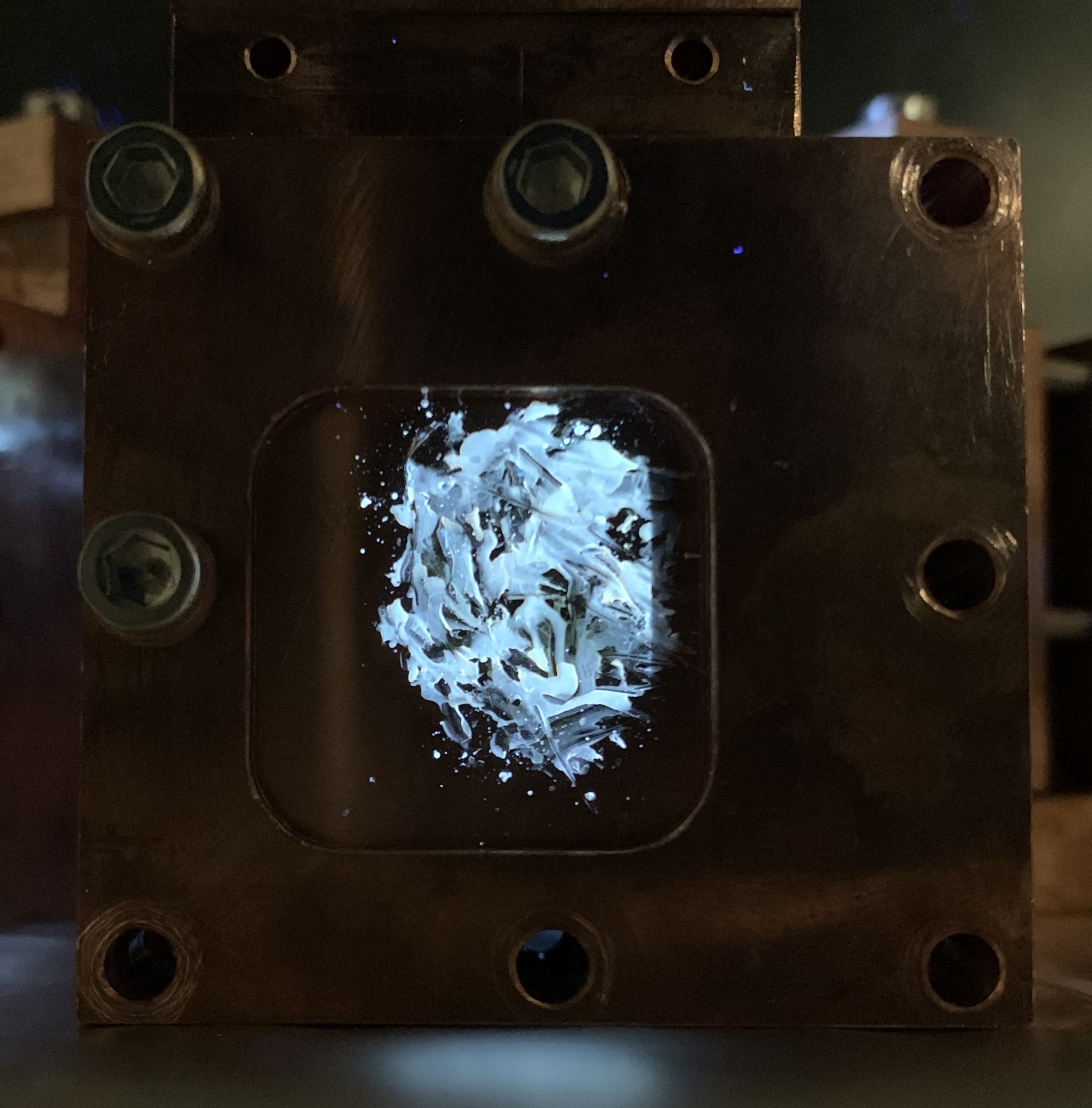}
 \\
 CW8400 &
 \scalebarimage{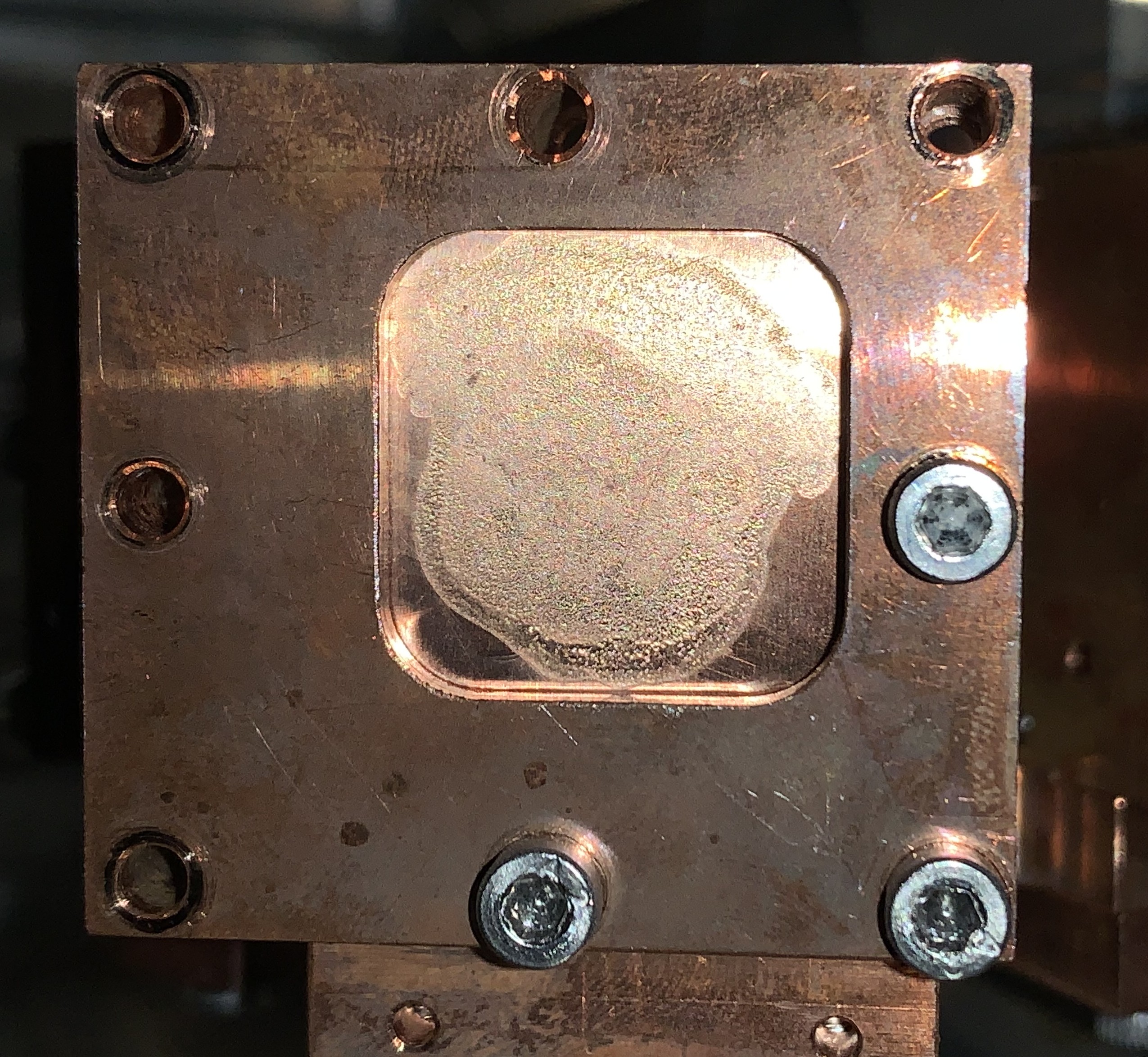}
 &
 \scalebarimage{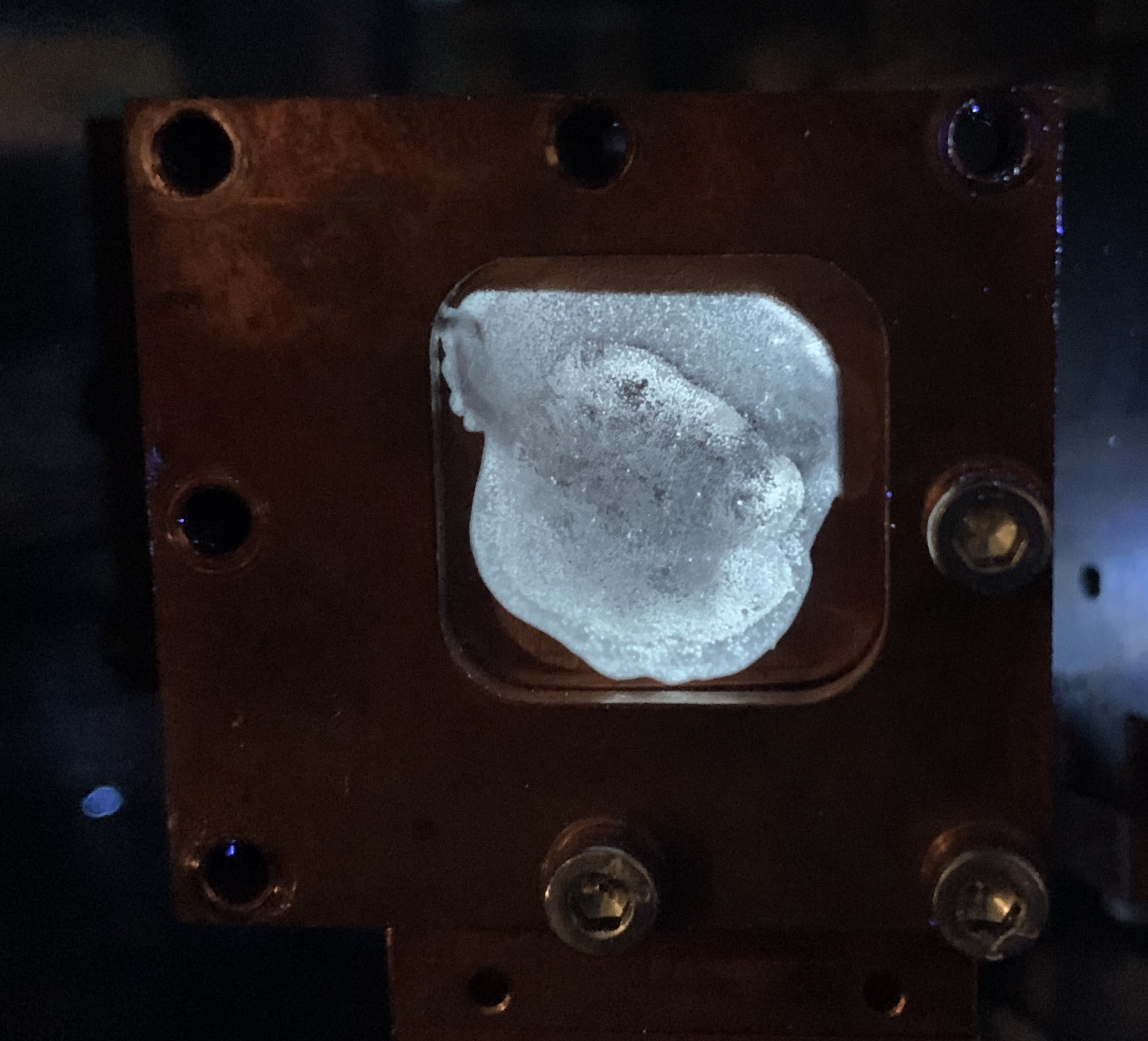}
 \\
 PEN &
 \scalebarimage{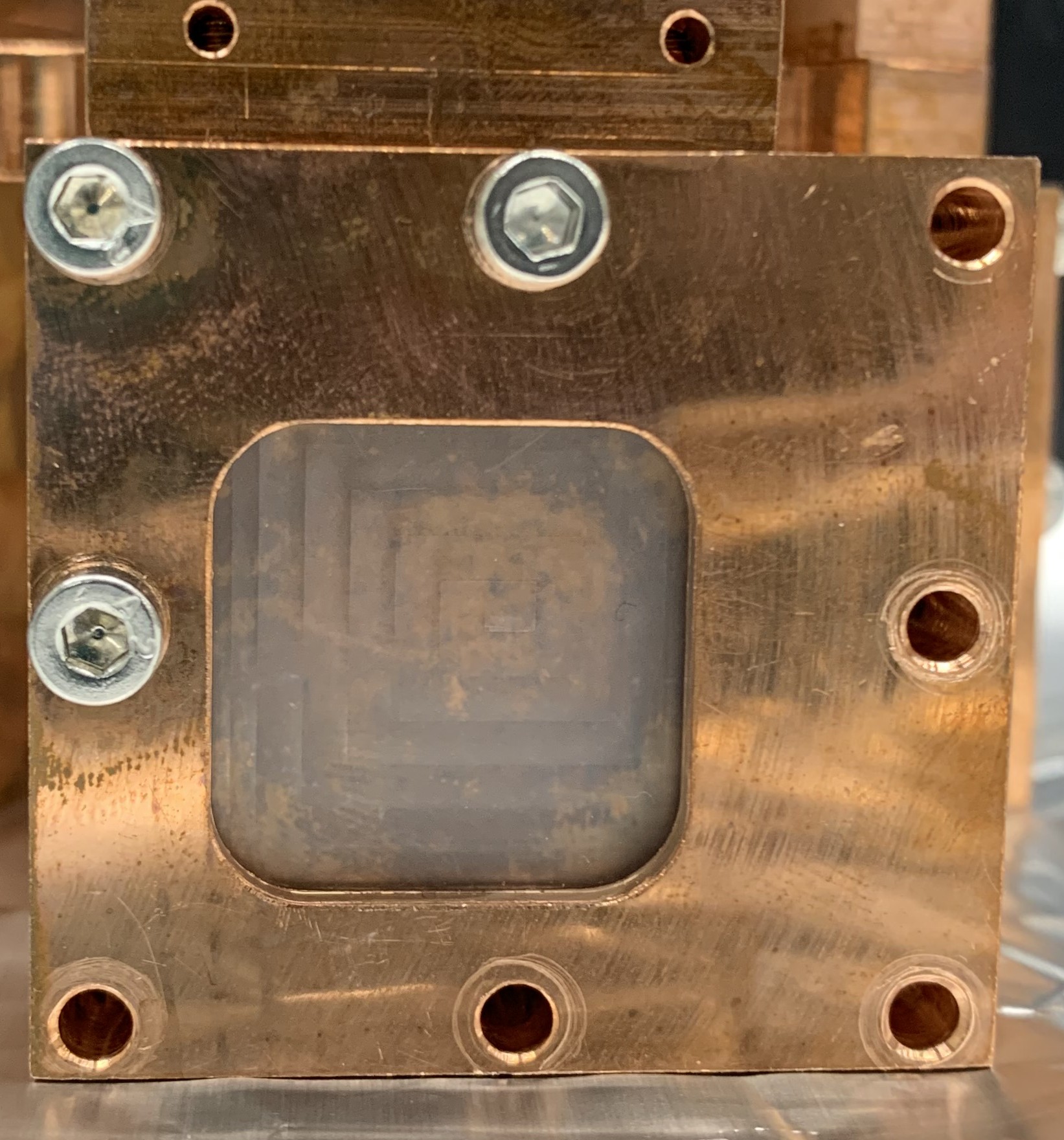}
 &
 \scalebarimage{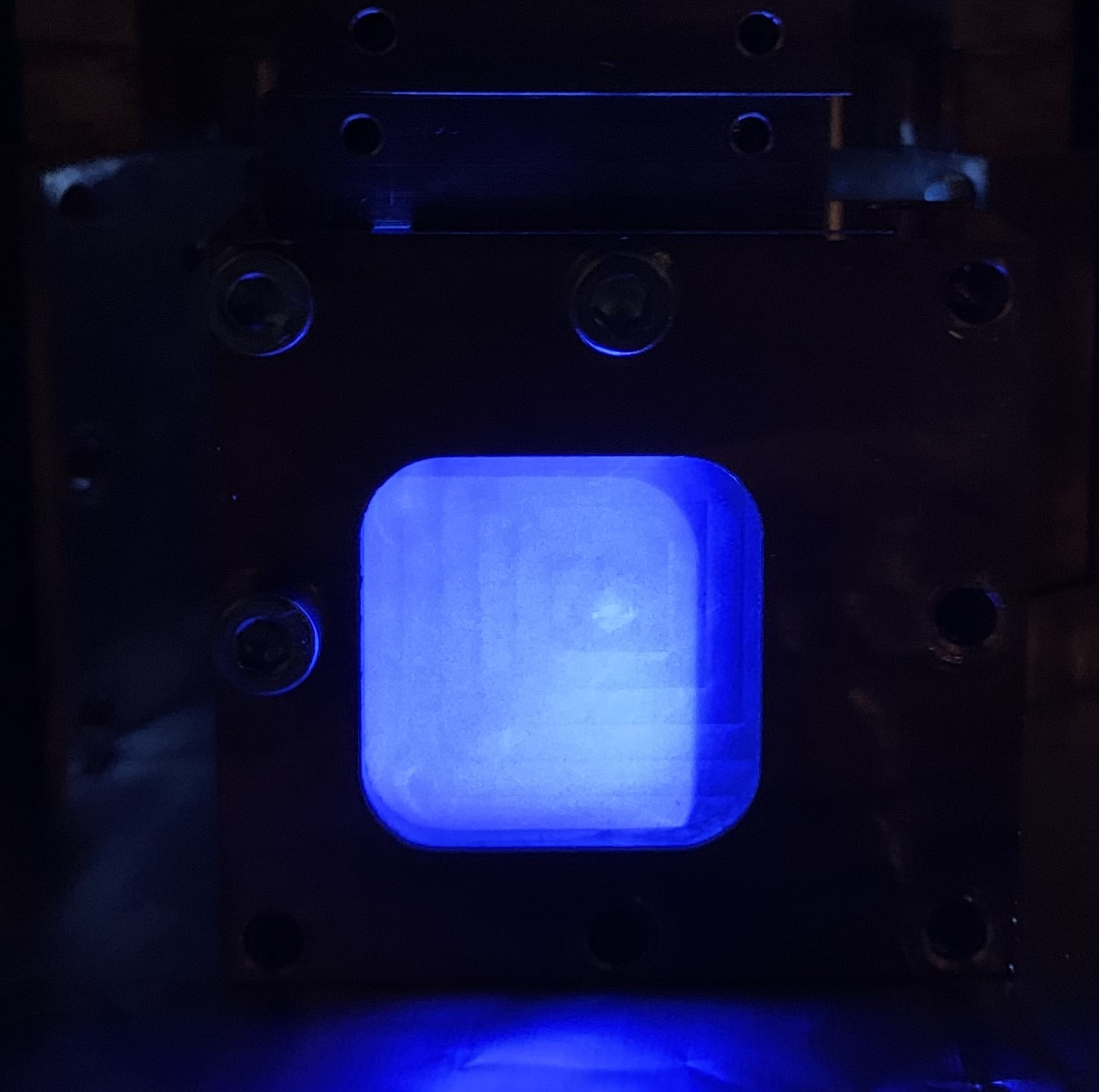}
 \\
\end{tabular}
\caption{
Photographs of the samples considered in this work are shown under white LED and under UV light (\SI{365}{\nano\meter}) while installed in the copper sample holder (compare Fig.~\ref{fig:setup} part d.).
}
\label{fig:sample_photos}
\end{figure}

\subsection{Emission spectra as a function of temperature under 310 nm excitation}
\subsubsection{Setup and procedure}
For measurements of PL emission spectra at different sample temperatures the setup described in detail in Ref.~\cite{Leonhardt_2024} was used. Samples were installed in a vacuum chamber (with pressures ranging from \SIrange{1e-4}{1e-8}{mbar} depending on the sample temperature) in a holder attached to a cold-finger. The samples were excited with \SI{310}{\nano\meter} UV-B photons from a Marktech Optoelectronics MTE310H33-UV LED. The PL light was collected with a lens and transmitted via optical fibre to an Ocean Optics QE65000 UV-IR spectrometer connected to a computer. Spectra were recorded with the OceanOptics-provided software. For each target temperature, six sequential emission spectra were recorded over a period of 2~minutes both during cool-down and during subsequent warm-up.

\subsubsection{Analysis and results}
\begin{figure}[!htpb]
    \centering

    \includegraphics[width=0.49\textwidth]{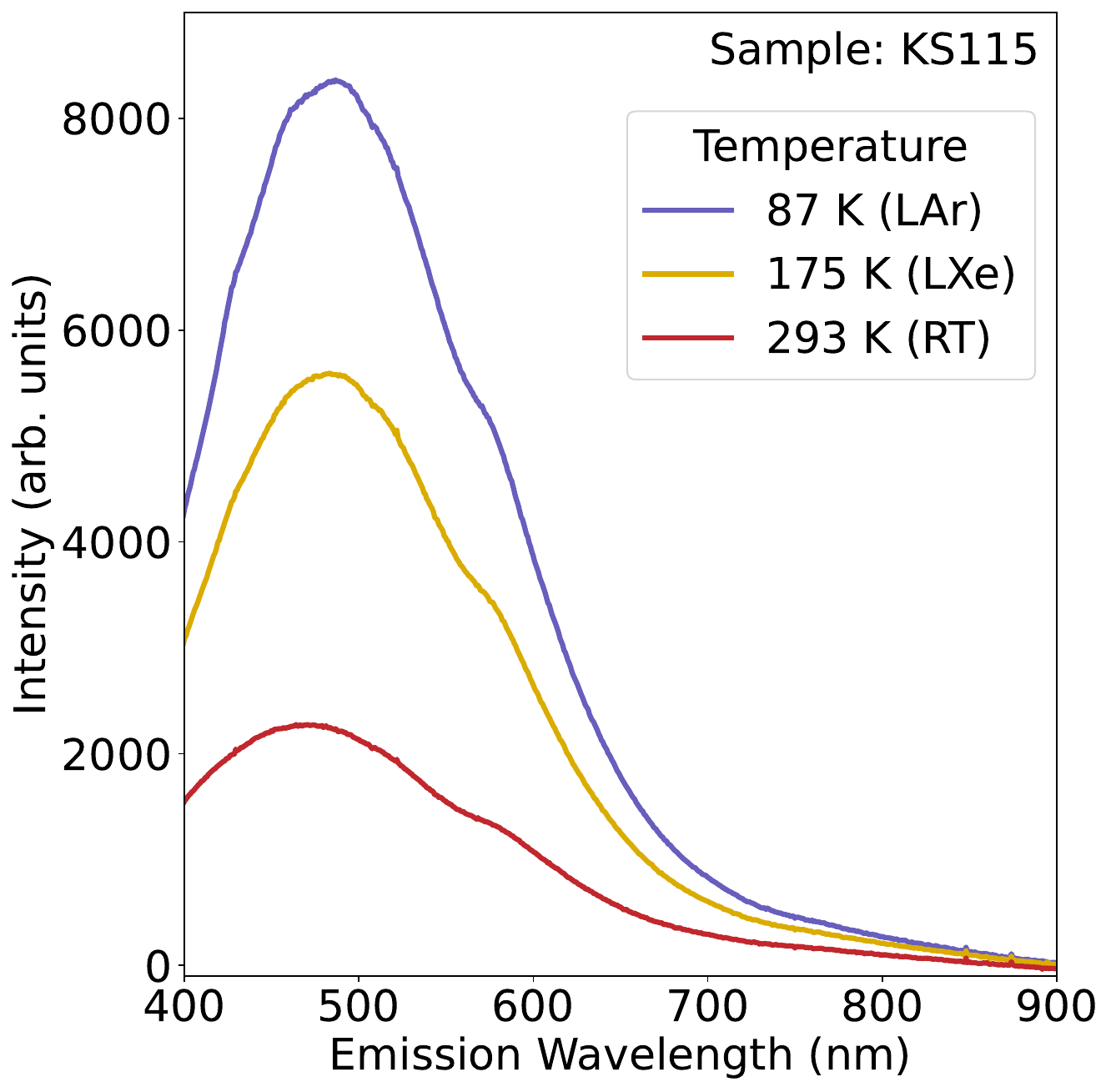}
    \includegraphics[width=0.47\textwidth]{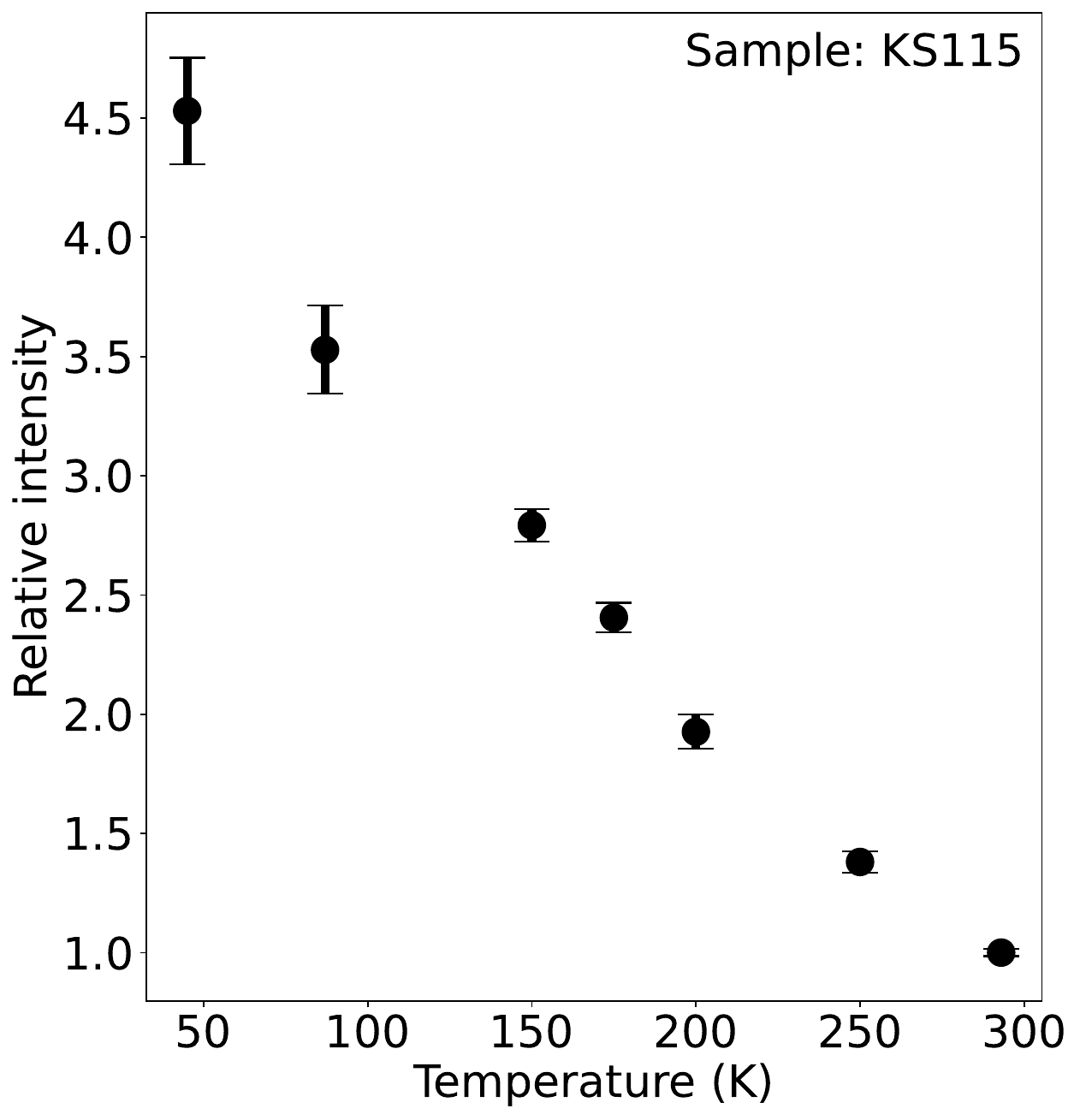}

    \vspace{0.2cm}

    \includegraphics[width=0.49\textwidth]{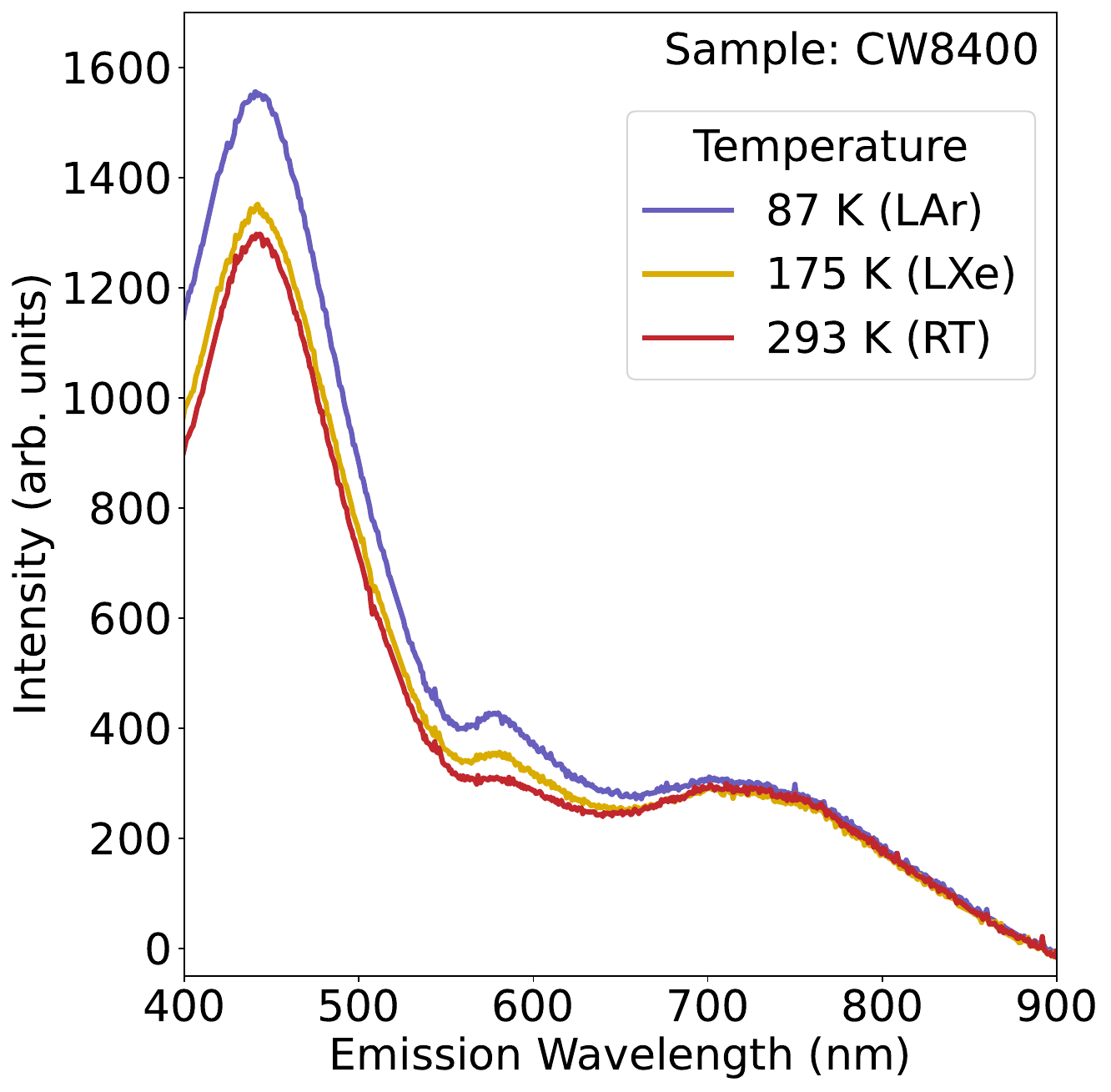}
    \includegraphics[width=0.48\textwidth]{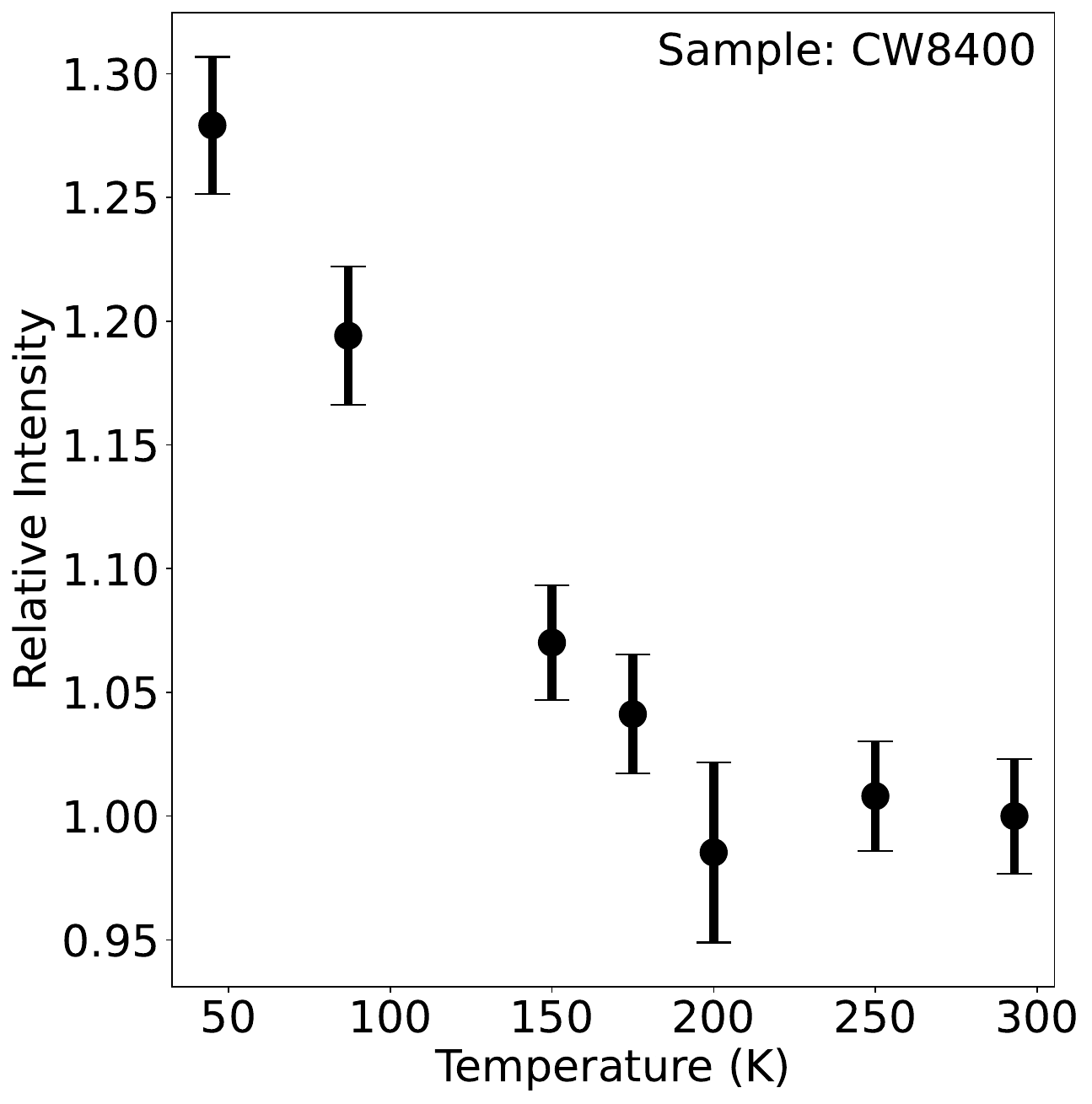}

    \caption{PL under \SI{310}{\nano\meter} excitation for the two flux samples. Left column: emission spectra at three relevant sample temperatures. Right column: integrated intensity in \SIrange{400}{900}{\nano\meter} range as a function of temperature with intensities normalised to the room temperature (\SI{293}{K}). The vertical error bars represent the stability of the setup conditions, calculated from the standard deviation of repeated measurements at each temperature.}
    \label{fig:emission_combined}
\end{figure}

Emission spectra recorded for the flux samples under \SI{310}{\nano\meter} excitation at LAr (\SI{87}{K}), LXe (\SI{175}{K}), and room temperature (\SI{293}{K}) are shown in Fig.~\ref{fig:emission_combined} (left). Data for wavelengths corresponding to 30 malfunctioning spectrometer channels (1.1\% of all channels) were removed, and dark counts were subtracted. The integrated intensities of the spectra from \SIrange{400}{900}{\nano\meter} as a function of sample temperature are shown in Fig.~\ref{fig:emission_combined} (right), relative to the room temperature measurement. The lower boundary of the integration window was set at the wavelength where the spectrum of the excitation light becomes negligible in intensity; the upper boundary corresponds to the limit of the spectrometer sensitivity. The error bars in Fig.~\ref{fig:emission_combined} represent the standard deviation across the repeated measurements to reflect instrumental repeatability. 

Like many PL materials, the flux residues studied exhibit higher PL efficiency at lower temperatures: at LXe temperature for example, the PL efficiency is 
$2.40\pm0.06$ times higher than at room temperature for KS115 flux residue, and 
$1.04\pm0.02$ times higher for CW8400 flux residue.

\subsection{Emission intensity under VUV excitation}
\subsubsection{Setup and procedure} \label{subs:setup}

\begin{figure}[htb]
    \centering
    \includegraphics[width=\textwidth]{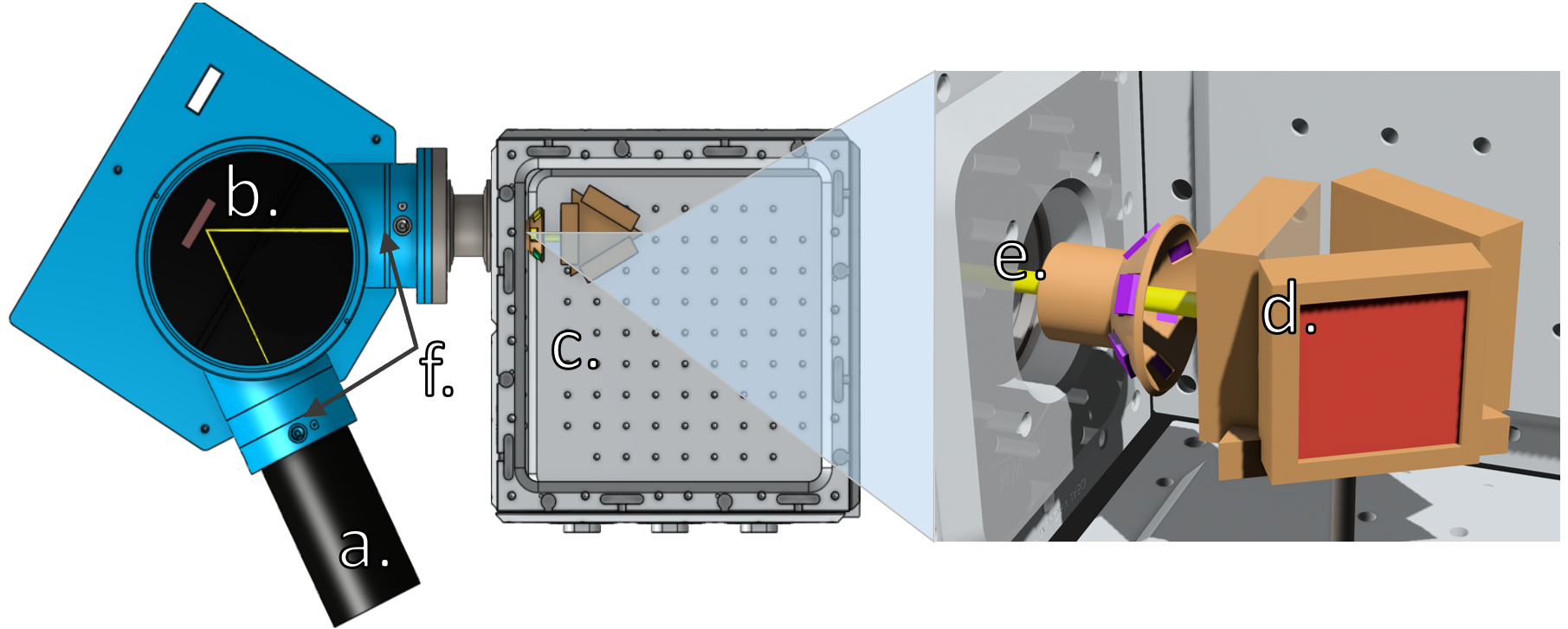}\hfill
    \caption{Top view of the experimental setup (left) and 3D zoom (right). Light is emitted by the deuterium lamp (a). It enters the monochromator (b) through the entrance slit and gets diffracted by the grating, then leaves the monochromator through the exit slit and enters the sample chamber (c). The sample holder (d) holds up to three samples and can rotate around a central vertical axis; the holder is made from copper (brown in the figure) and the samples are shown in red. Photoluminescence from the sample is recorded by six SiPMs mounted in the sensor holder (e). The entrance and exit slits of the monochromator are indicated by the arrows (f).}
    \label{fig:setup}
\end{figure}

Fig.~\ref{fig:setup} shows a sketch of the setup optimised to detect low-efficiency PL. A high-intensity Hamamatsu L15094 H$_2$D$_2$ lamp~\cite{lamp} is coupled to the entrance slit of a McPherson model 234/302 monochromator with \SI{1200}{g/mm} Al+MgF$_2$ grating, whose exit slit is connected to an IdealVac 6$\times$12$\times$12 sample chamber. 

The entire system is maintained under vacuum at a pressure of approximately \SI{1e-5}{mbar}. The sample holder, located along the beam axis inside the chamber, can rotate about its central vertical axis to expose one of three samples to the incident light. The beam spot on the sample is rectangular and has a size of approximately \SI{6}{\milli\meter} by \SI{4}{\milli\meter}. The samples are viewed by six Hamamatsu S13360-6050CS silicon photomultipliers (SiPMs)~\cite{sipms}, arranged around a hollow cone surrounding the beam (see Fig.~\ref{fig:setup}). Assuming uniform emission, the SiPMs are hit by 7\% of the PL photons emitted into the hemisphere from which the incident light comes. The SiPMs are sensitive in the wavelength range \SIrange{280}{900}{\nano\meter}, with a peak sensitivity at \SI{460}{\nano\meter}, and therefore are (nominally) insensitive to the VUV excitation light~\cite{sipms}. The wavelength detection efficiencies of these SiPMs weighted by the sample emission spectra shown in Fig.~\ref{fig:emission_combined} and Ref.~\cite{Leonhardt_2024} are approximately 32\% for KS115 solder flux, 29\% for CW8400 solder flux and 39\% for PEN. 

Due to the low light levels involved, PL intensity is determined by measuring the rate of individually detected PL photons and the SiPMs are therefore operated in photon-counting mode. The lamp provides steady-state illumination of the samples. The SiPMs are biased and read out through approximately \SI{30}{cm} long kapton-clad wires connected to a D-Sub vacuum feedthrough. On the air side, the signals are processed by a custom-designed readout circuit with an amplification factor of $80$ and a low-pass filter with a cutoff frequency of \SI{12}{MHz}. The amplified signals are digitised using a CAEN V1730 14-bit, \SI{500}{MS/s} waveform digitiser. Data were acquired for \SI{30}{\micro\second}-long windows at a repetition rate of \SI{30}{Hz}, set by an external pulse generator. Individual photon pulses are identified in each trace and counted, with the accumulated photon count used to determine the fluorescence rate. 

Data were taken in the following configurations:
\begin{enumerate}
    \item The blank copper plate, reference PEN, and one flux sample mounted in the sample holder \label{item:standard}(standard configuration).
    \item The blank copper plate and the two flux samples mounted in the sample holder\label{item:twosamples} (for a direct comparison between the flux samples). 
    \item A S13360-6050CS VUV-blind SiPM mounted in the sample holder (as a sample) and thus directly exposed to the beam \label{item:sipmpl} (to verify the response of this SiPM model to VUV light).
    \item A VUV-sensitive SiPM (Hamamatsu S13370-3050CN~\cite{sipmsvuv}) mounted in the sample holder and thus directly exposed to the beam \label{item:lampspectrum} (to measure the lamp spectrum).  
\end{enumerate}

In each configuration, once the required vacuum level was reached, `dark' data were taken with the lamp off. The lamp was then turned on and left for at least 5~minutes to stabilise before data were taken for each of the samples in the holder. The relatively bright PEN sample was measured with both monochromator slits open to just \SI{0.2}{mm}. For all other samples, the slits were opened to \SI{3.0}{mm} (maximum opening). These values were chosen to keep the signals within the linear dynamic range of the SiPMs. The slit sizes and the desired excitation wavelength were manually adjusted with the dials on the monochromator and the data were recorded for at least \SI{30}{\milli\second} per excitation wavelength, long enough to make statistical uncertainties on the count rate negligible. The lamp was then turned off and another `dark' dataset was recorded to track possible changes in the dark count rate.

\subsubsection{Analysis}\label{sect:analysis} 

\begin{figure}[htb]
    \centering
    \includegraphics[width=\textwidth]{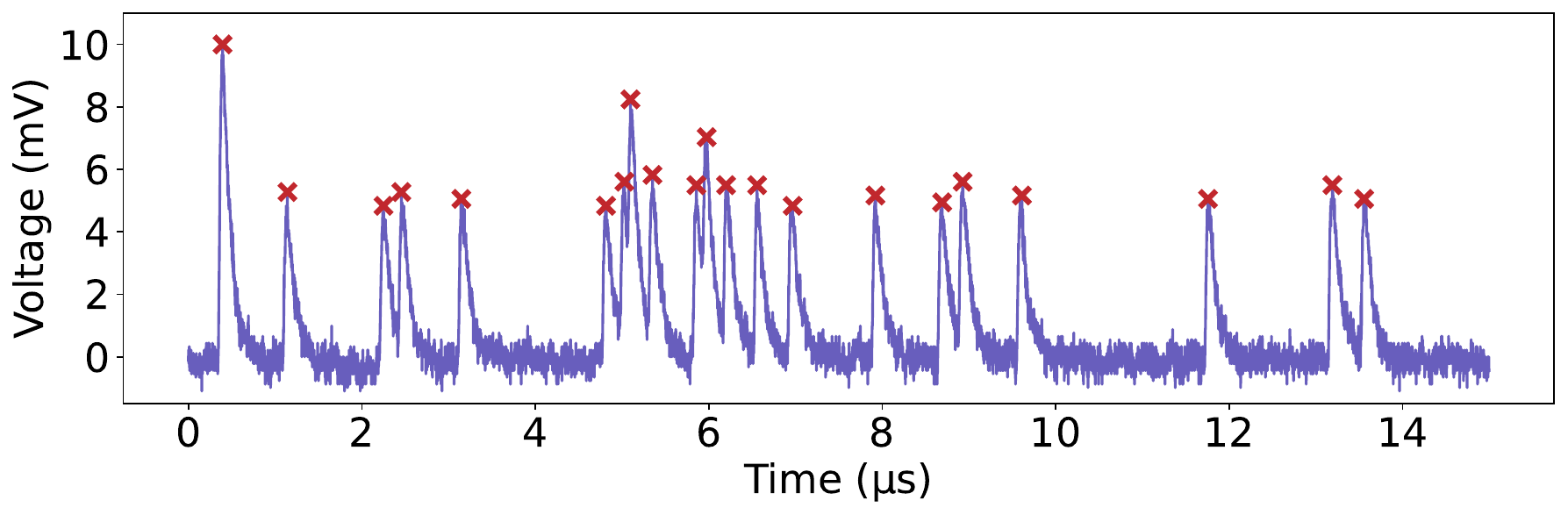}
    \caption{Baseline-corrected digitised SiPM trace. Peaks found by the \texttt{find\_peaks} algorithm are indicated by crosses and correspond to 1 or 2 photoelectrons.}
    \label{fig:waveform}
\end{figure}

A baseline-corrected digitised trace from a single SiPM is shown in Fig.~\ref{fig:waveform}. The trace contains multiple peaks, most of which have areas consistent with a single photoelectron. Peak times are extracted using SciPy’s \texttt{find\_peaks} algorithm~\cite{scipy_2020}. 
For each peak, the intersections of the rising and falling edges with the baseline are identified and the signal area between these two points is integrated. If the integration window contains overlapping pulses, the enclosed area is counted only once. The gain of each SiPM was determined for each run following the procedure in~\cite{sipm_manual}. The total integrated area of all peaks in the trace is divided by the gain to obtain the total number of detected photoelectrons. Dividing this by the measurement duration yields the photoelectron count rate. The count rate includes both real photons and instrumental effects (dark counts). The dark count rate is obtained from the lamp-off measurement and is subtracted from the rates measured with the lamp on to obtain the photodetection rate.
In case of statistically significant differences between the dark count rates measured at the beginning and end of a measurement campaign, the average of the two values was used, and uncertainties on the photodetection rate were increased accordingly.

The relative intensity of the excitation light from the lamp at different slit sizes and over time, as well as the reproducibility of the slit settings, was determined using configuration \ref{item:lampspectrum} (see Sect.~\ref{subs:setup}). Statistical uncertainties from Poisson counting statistics are negligible. Systematic uncertainties were estimated by accounting for fluctuations in the dark count rate, lamp intensity, and slit size.

Although the peak excitation wavelength reproducibility and calibration of the monochromator are both sub-nm, and therefore negligible for the present analysis, the spectral bandwidth incident on the samples depends on the monochromator exit-slit width and the diffraction-grating dispersion. For the exit slit openings used in this work, the transmitted spectrum has a full width at half maximum of \SI{0.3}{\nano\meter} at the smallest and \SI{10}{\nano\meter} at the largest slit size used; the latter is comparable to the widths of the LAr and LXe scintillation spectra.

\subsubsection{Results}

\begin{figure}[H]
    \centering
    \includegraphics[width=0.8\textwidth]{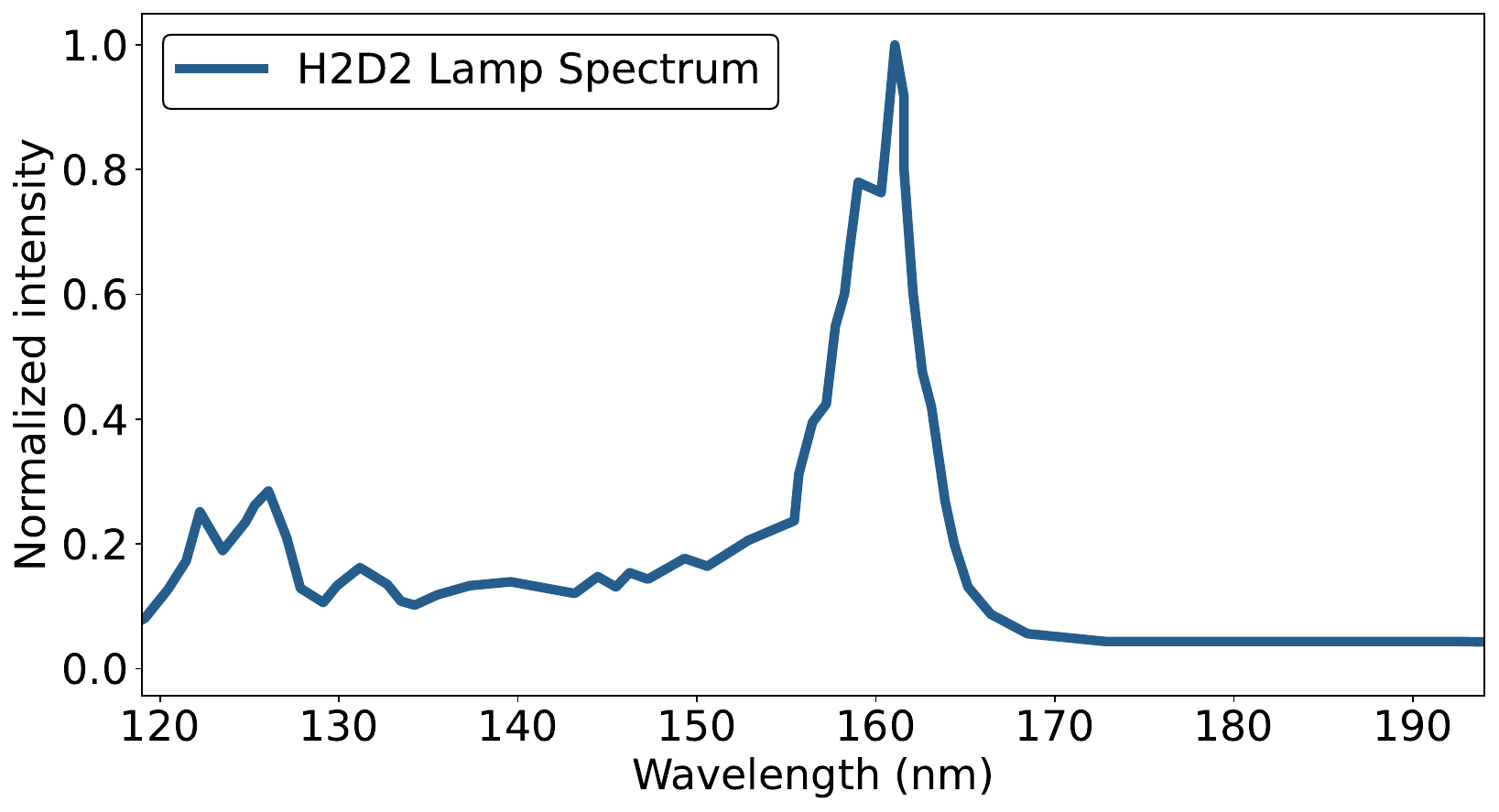}
    \includegraphics[width=0.8\textwidth]{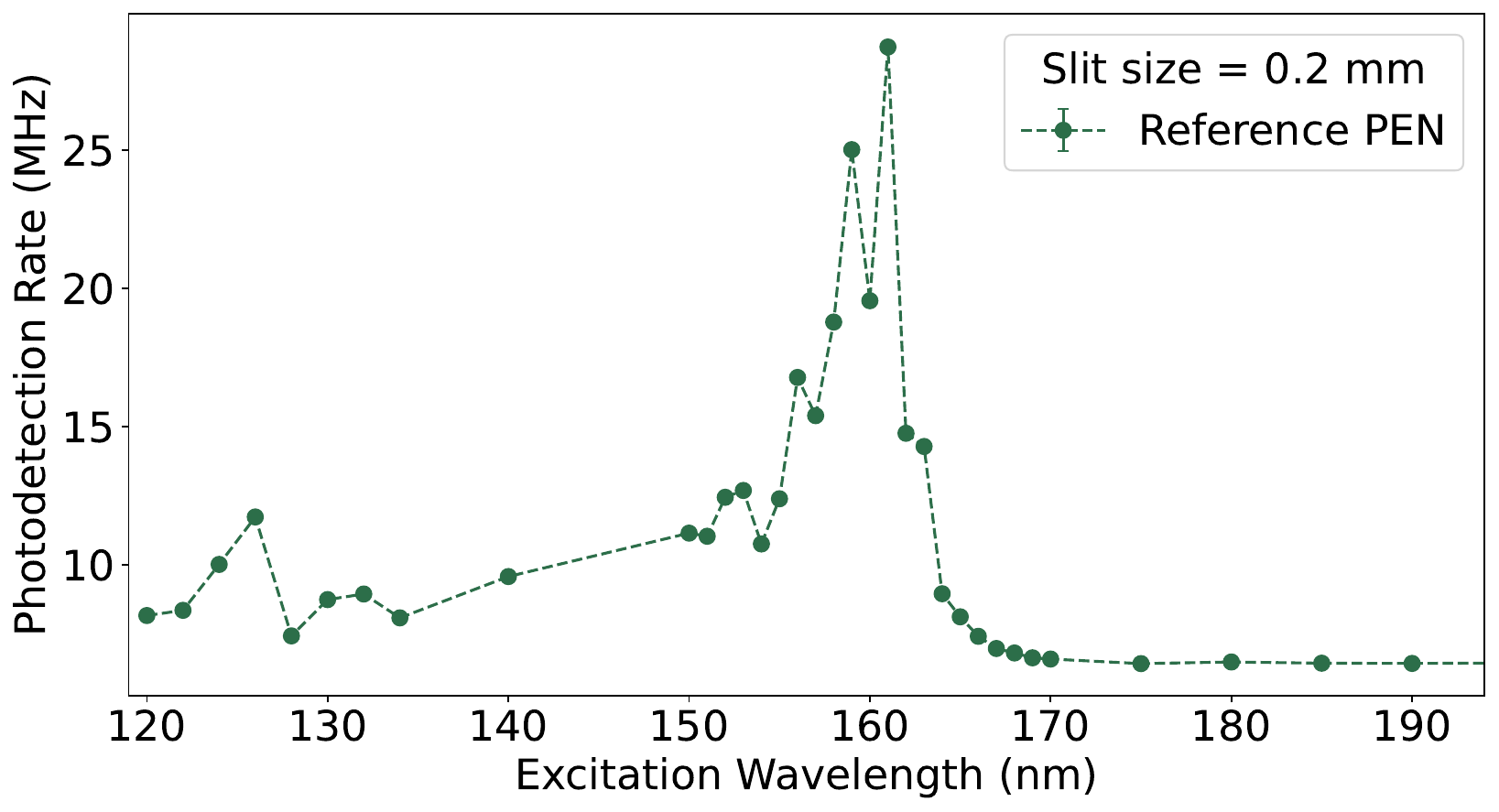}
    \includegraphics[width=0.8\textwidth]{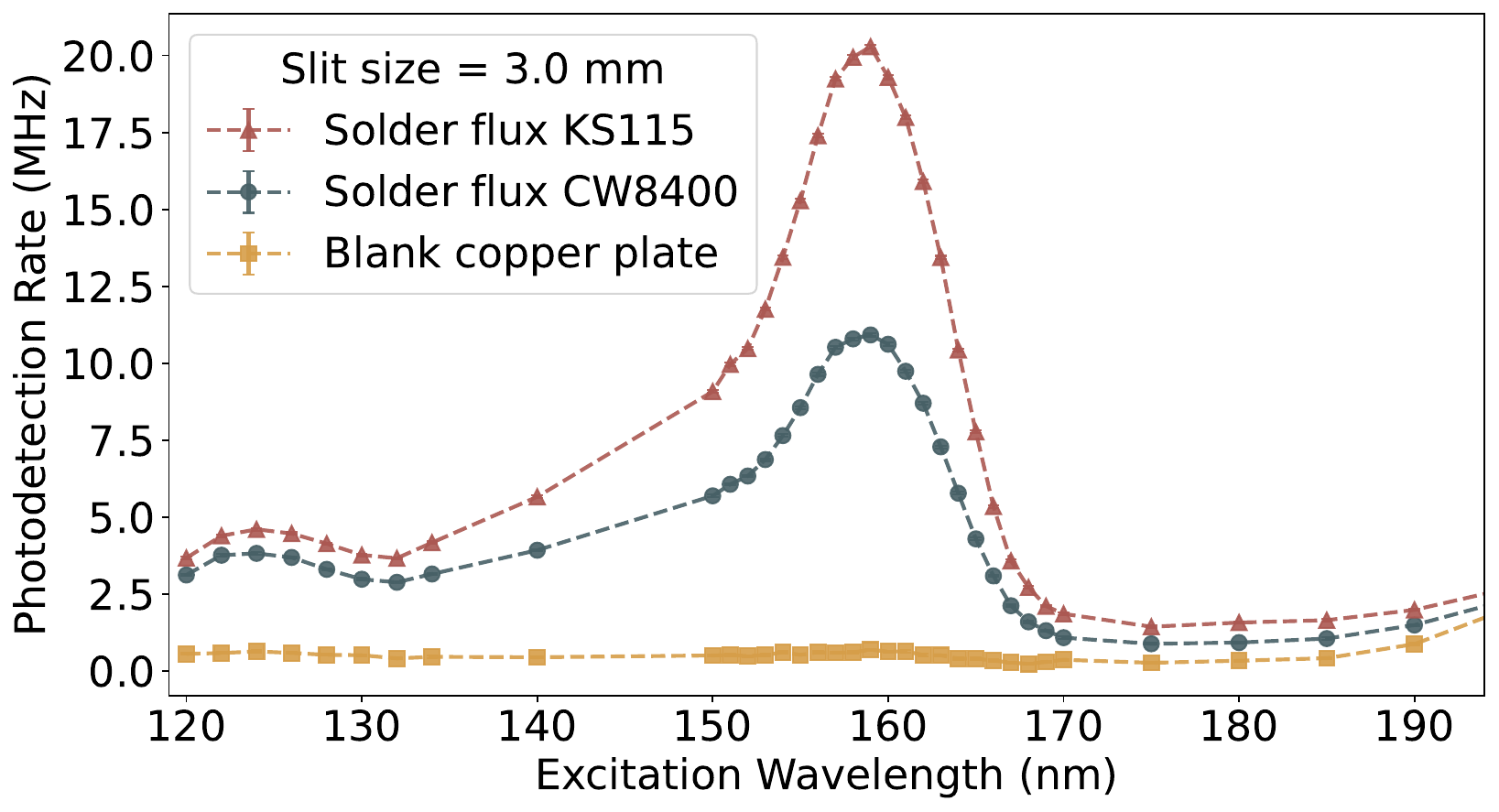}
    \caption{Top: Normalised emission intensity of the H$_2$D$_2$ lamp measured in the wavelength range used for the sample illumination (\SIrange{120}{200}{\nano\meter}). Middle and bottom: PL emission intensity in the SiPM sensitivity range (\SIrange{280}{900}{\nano\meter}), for the reference, solder flux residue, and blank samples as a function of excitation wavelength. Vertical error bars are derived from Poisson uncertainty and are too small to be visible.
    }
\label{fig:emissionintensity}
\end{figure}

Figure~\ref{fig:emissionintensity} (top) shows the VUV spectrum of the lamp, measured with the VUV-sensitive SiPM directly exposed to the beam (configuration~\ref{item:lampspectrum} in Sect.~\ref{subs:setup}) and normalised to the intensity of the brightest peak. The measured spectrum is consistent with the lamp datasheet~\cite{lamp}. Figure~\ref{fig:emissionintensity} (middle) shows the photodetection rate from the PEN sample (configuration~\ref{item:standard} in Sect.~\ref{subs:setup}), recorded with the same slit size as the spectrum above it, and following the excitation light spectrum closely. Figure~\ref{fig:emissionintensity} (bottom) shows the photodetection rate for the blank copper, the KS115, and the CW8400 flux residue samples (configuration~\ref{item:twosamples}) as a function of excitation wavelength. The response of the flux samples follows the lamp spectrum, clearly indicating the presence of photons emitted by the samples within the SiPM spectral sensitivity range (\SIrange{280}{900}{\nano\meter}), and thus demonstrating PL from the samples. The spectral shape is broader than the spectrum shown in the top panel because of the wider slit size increasing the bandwidth of the excitation light (see Sect.~\ref{sect:analysis}). By contrast, the response of the blank copper is significantly lower and is almost consistent with dark counts up to approximately \SI{180}{\nano\meter}.

\begin{figure}[h]
    \centering
    \includegraphics[width=0.8\textwidth]{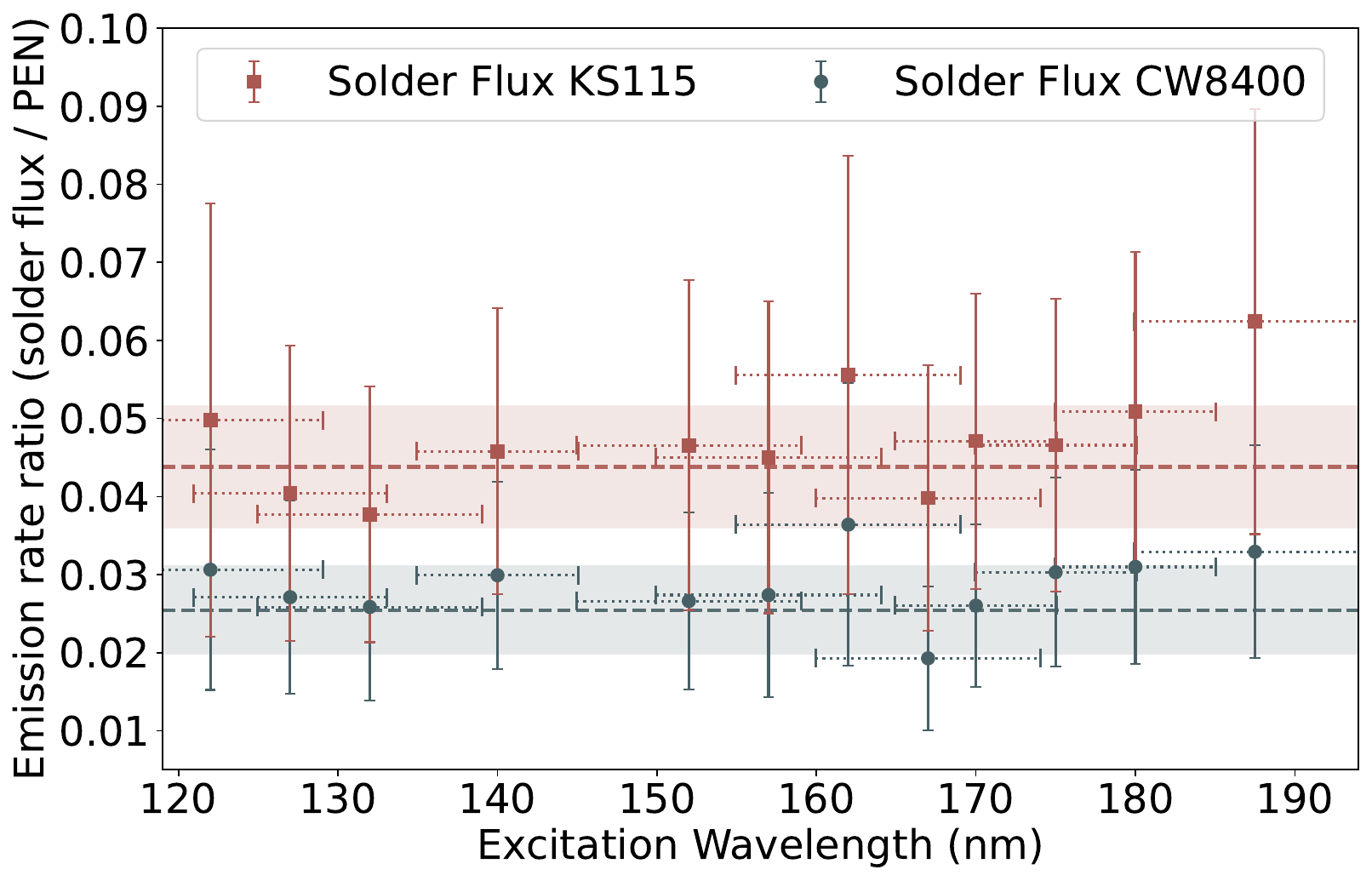}
    \caption{Relative PL efficiencies of the flux samples with respect to the PEN reference sample at room temperature. Measurements taken within \SI{5}{\nano\meter} bins on the x-axis are averaged into single points to improve visual clarity. The dashed line and the shaded band represent the weighted mean and the RMS of the unbinned data. The dominant component to the vertical error bars is due to instrumental corrections applied to the data. The dotted horizontal bars represent a range of excitation wavelength over which the data is integrated and averaged. }
    \label{fig:efficiency}
\end{figure}

For excitation wavelengths above \SI{180}{\nano\meter}, the photon detection rate increases for both the flux samples and the copper blank, which is inconsistent with the lamp spectrum. To investigate this effect, data were taken in configuration~\ref{item:sipmpl}.
These measurements show that the nominally VUV-blind 13360-6050CS SiPMs do have some sensitivity to VUV light and are increasingly sensitive above approximately \SI{180}{\nano\meter}. Inspection of the SiPM under UV-A (\SI{365}{\nano\meter}) illumination revealed white PL around the edges where the device is mounted to its backing. We therefore infer that the PL packaging shifts the VUV light in wavelength toward the SiPMs' nominal sensitivity range, and we attribute the rising count rates from the blank sample to reflected excitation light\footnote{We verified that Hamamatsu S13360-6050PE SiPMs, which have a different packaging, are also photoluminescent, and did not consider other models as they would not have had the right detection efficiency and dark count characteristics.}. We account for this effect by using asymmetric error bars in Fig.~\ref{fig:efficiency}: The central position of the point is determined in the same way for all wavelengths, and so is the size of the upward error bar. The size of the downward error bar takes into account that the flux samples may be at most as reflective to VUV light as the blank copper samples and that the count rate from the blank copper should therefore be subtracted from that of the samples. This is an over-estimate of the effect, as the copper is more reflective than the flux samples, so error bar lengths are conservative.

Fig.~\ref{fig:efficiency} shows the intensity of PL light detected from each of the two flux residues relative to the reference PEN sample. Across the VUV wavelength range, the average PL response relative to PEN is 2.5\% $\pm$ 0.6\% for CW8400 and 4.4\% $\pm$ 0.8\% for KS115, where uncertainties correspond to the RMS across measurements. These values include a correction for the SiPM efficiency over the emission spectrum of each sample, as well as corrections for the difference in excitation light intensity between measurements due to the difference in slit size. We observe no significant dependence of the response on wavelength, though this may be due to large systematic uncertainties related to correcting for different monochromator slit sizes.

Assuming the temperature dependence of the emission intensity is independent of excitation wavelength, this translates to approximately 2\% for CW8400 and 8\% for KS115 with respect to PEN at LXe temperature. These values were obtained by using the increase in emission intensity for LXe temperature (\SI{175}{K}) from Sect.~\ref{sec:spectrofluorometry} for the two fluxes while the value for PEN was extrapolated from Fig.~4 in Ref.~\cite{Leonhardt_2024}. We estimate that the absolute PL efficiency of the PEN reference sample is between approximately 50\% and 80\% at LXe temperature \cite{Leonhardt_2024, araujoWavelengthshiftingReflectorsCharacterization2022}. 

\section{Discussion and conclusion}\label{Sec:Conclusion}

\begin{figure}[h]
    \centering
    \includegraphics[width=0.5\textwidth]{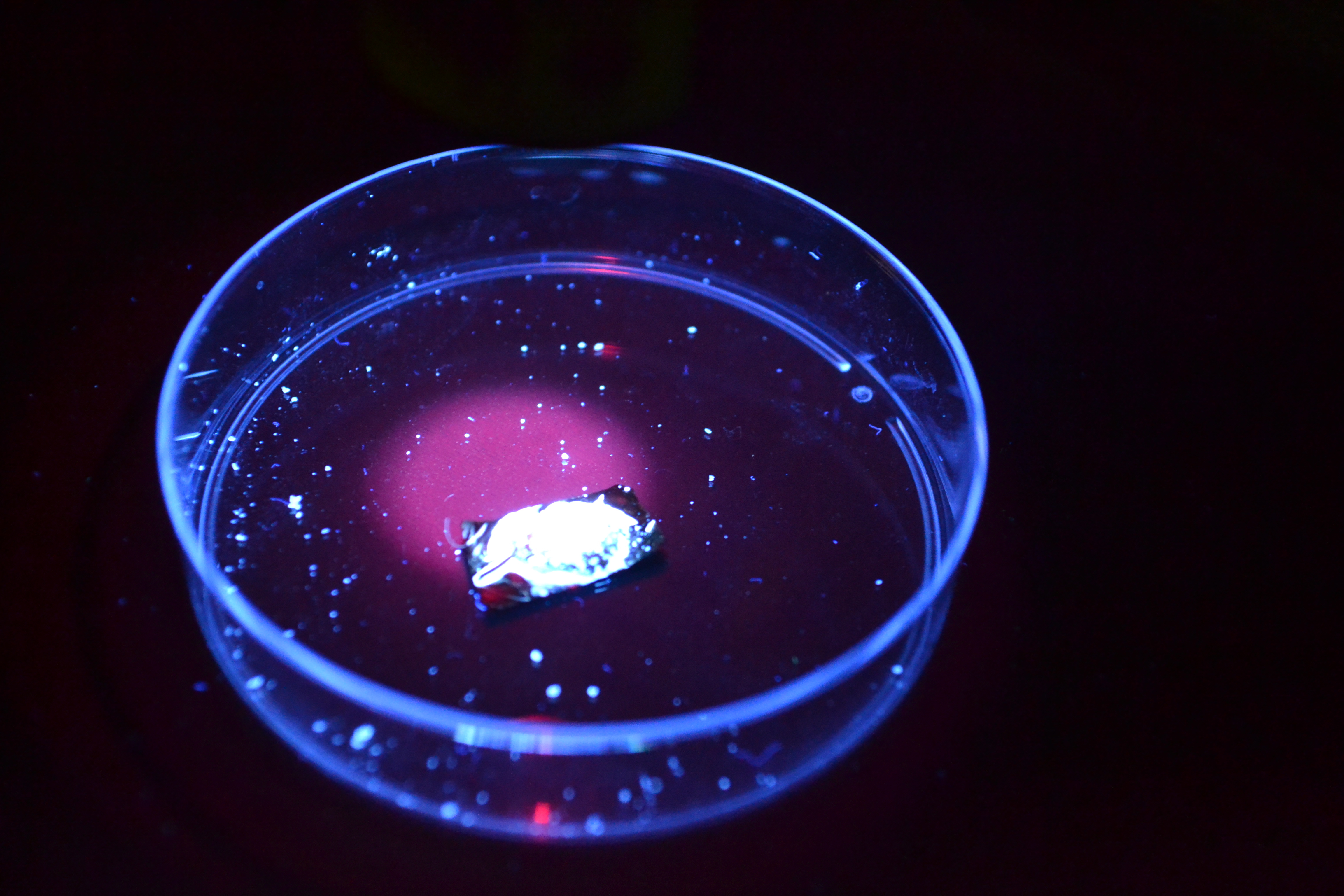} 
    \caption{A solder point of KS115 solder flux on a small piece of copper (centre of the Petri dish) after it was cooled to approximately \SI{80}{K}, shown illuminated with \SI{365}{\nano\meter} light. Flux residue PL appears blue-white. Many small photoluminescent flakes are visible in addition to the flux residue covering the solder point. The flakes were not there before cooling the sample, and the shape of the bigger flakes matches gaps in the flux residue covering the solder point. This suggests that thermal stress caused the small pieces of the solder flux to break loose.}
    \label{fig:stress_result}
\end{figure}

We have shown that solder fluxes commonly used in dark matter detectors exhibit photoluminescence in response to LAr and LXe scintillation light, with PL efficiencies of order 1\%.  Although solder joints are typically located outside the active detector volume, PL photons from flux residue may nonetheless contribute to unexpected signals if they are visible from the active region. Several mechanisms could enable this: dissolution of flux residue in the noble liquid~\cite{rentzepis_xenon_1981, sponsler_liquid_1989} followed by re-deposition on surfaces facing the active volume\cite{shaw_ultraviolet_2007}; mechanical fragmentation of flux residue during thermal cycling (Fig.~\ref{fig:stress_result}), with small particles being carried by liquid flow into the active volume; and transmission of visible PL photons produced at solder joints through the (usually PTFE) reflectors that separate the active volume from the inactive volume. 

Solder flux residues can release electronegative impurities into the liquid, which could lead to a reduction in the free-electron lifetime and thereby impact the charge extraction from the TPC\cite{baudis_electron_2023}. However, the magnitude of this effect depends on the composition and amount of residue, as well as on the effectiveness of the purification system and on the timescales over which these impurities are released. Therefore, a free-electron lifetime large enough for successful long-term TPC operation does not exclude the presence of photoluminescent solder flux residues.

Prompt PL generated inside the active volume could bias the reconstructed event energy and/or position, while emission from long-lived PL states could contribute to event-time-correlated instrumental backgrounds\cite{singh_search_2025}. PL generated outside the active volume and shining through the PTFE reflectors produces a signal without a corresponding particle interaction in the active volume, leading to time-uncorrelated background.

The measured PL efficiencies have large systematic uncertainties. The measurements rely on comparisons to a reference sample with different geometry and optical properties, and no corrections are applied for differences in light transport, emission angle, or uniformity of the samples. In addition, emission spectra and temperature dependence were measured under near-UV excitation, while VUV excitation measurements were performed at room temperature, and we assume that these properties do not depend on the excitation wavelength.

Despite these limitations, this study provides the first experimental evidence that solder flux residues can photoluminesce under VUV excitation at levels potentially relevant to noble-liquid dark matter detectors. These results motivate further dedicated measurements, in particular of photoluminescence efficiency and time profiles under simultaneous VUV excitation and cryogenic conditions, to better assess the possible impact on detector backgrounds.

\acknowledgments
\vspace{-0.5cm}
We would like to thank Joost van Dijk and Auke-Pieter Colijn for valuable discussions and feedback, and Julio Acosta, Martijn van Overbeek and the LNGS workshop for their invaluable practical support. Parts of this work were supported by the Volkert van der Willigen grant and by the DFG through the Excellence Cluster ORIGINS EXC 2094 — 390783311 and the SFB1258. 

\vspace{-0.3cm}
\bibliography{solder_bibliography}
\bibliographystyle{sn-mathphys-num}
\end{document}